\pdfoutput=1

\documentclass[12pt]{article}

\usepackage{amssymb, amsmath, amsthm, mathtools, bm}

\usepackage{graphicx, xcolor}

\usepackage{booktabs, array, float, multirow}

\usepackage[T1]{fontenc}
\usepackage[utf8]{inputenc}
\usepackage{fullpage}
\usepackage{enumitem}
\usepackage[round]{natbib}
\usepackage[unicode,hidelinks]{hyperref}

\DeclareMathOperator{\Var}{Var}
\DeclareMathOperator{\E}{E}
\DeclareMathOperator{\CV}{CV}
\DeclareMathOperator{\logit}{logit}

\newcommand{\T}{\top}
\newcommand{\bx}{\bm{x}}
\newcommand{\bz}{\bm{z}}
\newcommand{\bh}{\bm{h}}

\newcommand{\bX}{\bm{X}}
\newcommand{\bbeta}{\bm{\beta}}
\newcommand{\bgamma}{\bm{\gamma}}

\newenvironment{keywords}{%
  \par\noindent\textbf{Key words:}\ }{%
  \par\medskip}

\begin{document}

\title{\textbf{Diagnostics-guided variance-inflated Fay--Herriot estimation from non-probability samples}}

\author{
Andrius~Čiginas\\[0.4em]
{\small Vilnius University, Faculty of Mathematics and Informatics,}\\
{\small Institute of Data Science and Digital Technologies}\\
{\small Akademijos str.~4, LT-08412 Vilnius, Lithuania}
}

\date{}

\maketitle

\begin{abstract}
Non-probability data sources are increasingly considered in small area estimation, but inverse probability weighting (IPW) gives model-dependent domain estimators whose reliability may vary substantially across domains. Standard Fay--Herriot (FH) smoothing borrows strength across domains, yet it uses the supplied area-level variance estimates as if they fully described the uncertainty of the input estimators. This can be misleading when some domains have weak coverage, unstable weights, or poor auxiliary balance, since these features may indicate selection-bias risk not captured by the estimated variance alone. We propose a diagnostics-guided variance-inflated FH estimator for finite-population domain totals. The method starts from calibrated IPW domain estimators, summarizes their reliability through a small set of domain diagnostics, and introduces a mixture variance-inflation component in the FH observation equation. Domains whose diagnostics indicate weaker IPW information are thereby smoothed more strongly toward the area-level regression mean. A truth-known validation based on a pseudo-real population of Lithuanian business enterprises shows a substantial reduction in estimation error relative to calibrated IPW.
\end{abstract}


\section{Introduction}\label{sec:introduction}

Non-probability data sources are increasingly considered in survey statistics and official statistics. Administrative records, business registers, platform data, web-scraped data, and other operational sources may contain many units and may provide either study variables or covariates related to them. However, membership in such sources is typically not generated by a probability sampling design. When they are used for finite-population inference, estimation must rely on modeling or adjustment assumptions, and the resulting uncertainty must reflect more than ordinary sampling variation. This issue is particularly important for domain and small area estimation, because the quality of a non-probability estimator may vary substantially across domains.

The literature relevant to the latter problem can be organized into four strands. The first strand develops general inferential frameworks for non-probability samples and for the integration of probability and non-probability data. Model-based adjustment, propensity-score weighting, calibration, mass imputation, and doubly robust estimation have been studied as ways to reduce selection bias when the non-probability source is supported by auxiliary information from a probability sample, a census, or a register \citep{ElliottValliant2017,ChenLiWu2020,YangKim2020,Wu2022}. In particular, data-integration methods allow the study variable to be observed only in the non-probability or big-data source, while common auxiliary variables are observed in a probability sample or in population data \citep{KimTam2021,YangKimHwang2021}. This literature provides the inferential background for inverse probability weighting (IPW), calibration, and adjusted use of non-probability data, but it is usually not formulated as an area-level small area model.

The second strand is closer to the present paper because it concerns domains or small areas. Big-data-assisted small area estimation has used large non-traditional sources as auxiliary information in model-based estimators \citep{MarchettiEtAl2015}. More recent work considers the case where a non-probability big-data source contains the study variable and a reference probability sample supplies auxiliary information, leading to doubly robust small area estimators that combine propensity weighting and prediction modeling \citep{SchirripaSpagnoloEtAl2025}. Related work studies domain estimation from weighted non-probability samples and emphasizes that pseudo-weighting may reduce selection bias but may also introduce additional variability and domain-level instability \citep{LiuEtAl2026}. Bayesian predictive small area inference under selection bias has also been proposed for small area proportions \citep{ChoiNandramKim2021}. These papers are directly relevant to the present work because they address domain or small area inference with non-probability data. However, they do not consider the area-level Fay--Herriot (FH) setting studied here, where the reliability of a calibrated IPW domain estimator is represented through diagnostics-guided inflation of the sampling-error variance.

The third strand combines probability and non-probability sources at an aggregated or model-based level. General discussions of data integration in official statistics emphasize the role of probability samples as anchors for inference from other sources \citep{Rao2021,Lohr2021}. Aggregated-level combinations of probability and non-probability estimates have been studied using bias and mean squared error models \citep{VillalobosAlisteEtAl2025}. A related two-step approach uses incomplete auxiliary information of non-probability origin in model calibration before a subsequent FH area-level model \citep{SlevinskasEtAl2026}. Work on unknown overlaps and thresholding of non-probability units develops Bayesian and pseudo-weighting strategies for combining probability and convenience samples and for improving domain estimation by reducing poor-support contributions \citep{SavitskyEtAl2023,SavitskyEtAl2025}. These approaches are useful for understanding bias--variance trade-offs in data integration. The setting considered in this paper is different: the study variable is not observed in a probability sample, and therefore, there is no direct probability-sample benchmark for estimating a domain-specific bias in the non-probability estimator.

The fourth strand is robust small area estimation and mixture modeling for FH-type models. Robust empirical best linear unbiased prediction, robust $M$-quantile, and related methods address outlying units or outlying area effects \citep{SinhaRao2009,ChambersEtAl2014,JiangRao2020}. Mixture alternatives to the FH model have also been used to accommodate uncertain or contaminated random effects \citep{DattaMandal2015,ChakrabortyDattaMandal2016}. These methods are important for positioning the present contribution, but their robustness target is not the same. They typically protect against model outliers or unusual area effects. In contrast, the variance inflation proposed here is attached to the observation equation and is driven by diagnostics of the non-probability IPW input, such as domain coverage, weight variability, and auxiliary balance.

We develop a diagnostics-guided variance-inflated FH estimator for finite-population domain totals from non-probability samples. In one domain, a calibrated IPW estimator may be supported by high coverage, stable weights, and good balance with respect to known auxiliary information. In another domain, the observed units may cover only a small part of the population, the calibrated weights may be highly variable, and the weighted sample may remain poorly aligned with the population. A classical FH model \citep{FayHerriot1979,RaoMolina2015} can smooth area-level inputs using domain covariates, but it uses the supplied area-level variance estimates as if they fully described the uncertainty of the input estimators. This can be misleading when diagnostic evidence indicates selection-bias risk or weak support that is not captured by the estimated variance alone.

The proposed approach has three components. First, domain totals are estimated by calibrated IPW. The calibration is formulated at the domain level and may use only the domain population size or may additionally use known auxiliary totals. Second, a small set of domain diagnostics summarizes the quality of the IPW input: domain coverage, calibrated weight variability, and auxiliary balance. These diagnostics are combined into a reliability score and used to define a set of relatively reliable domains. Third, the FH observation variance is inflated through a mixture mechanism for domains whose diagnostics indicate weaker information. The resulting estimator remains an area-level model estimator, but it is guided by evidence about the reliability of the non-probability input.

The empirical part is a truth-known validation based on a pseudo-real population of Lithuanian business enterprises. In this population, the target turnover variable is known for all units and is therefore available for evaluating actual estimation error, whereas fitting uses only the units belonging to the administrative non-probability source.

The contribution of the paper is threefold. We propose a compact diagnostics-based reliability construction for non-probability small-area input estimators; we incorporate that construction into a mixture variance-inflated FH estimator; and we validate the resulting estimator using truth-known business-turnover data. The method should not be interpreted as identifying arbitrary unmeasured selection bias without an external benchmark. Rather, it provides a transparent way to propagate diagnostic evidence about weak non-probability support into an area-level FH analysis.

The rest of the paper is organized as follows. Section~\ref{sec:setup} introduces the finite-population setup and calibrated IPW domain estimator. Section~\ref{sec:diagnostics} defines the domain diagnostics and reliability score. Section~\ref{sec:model} presents the diagnostics-guided mixture variance-inflated FH estimator. Section~\ref{sec:services} presents the business-turnover validation. Section~\ref{sec:discussion} concludes. Bootstrap details for the calibrated IPW variance input and for the full-pipeline mean squared error (MSE) estimator are given in Appendices~\ref{app:direct_bootstrap} and~\ref{app:full_bootstrap}.

\section{Basic setup and calibrated IPW estimation}\label{sec:setup}

Let
\begin{equation*}
U=\bigcup_{d=1}^D U_d,
\qquad
U_d\cap U_{d'}=\varnothing \quad (d\ne d')
\end{equation*}
be a finite population partitioned into $D$ domains. The domain size is $N_d=|U_d|$, and the target parameter is the finite-population domain total
\begin{equation*}
Y_d=\sum_{i\in U_d}y_i,
\qquad d=1,\ldots,D.
\end{equation*}
Let $S\subset U$ denote the observed non-probability sample, and define
\begin{equation*}
\delta_i=\mathbf 1(i\in S),
\qquad
S_d=S\cap U_d,
\qquad
n_d=|S_d|.
\end{equation*}
The values $y_i$ of the study variable are observed for $i\in S$. Auxiliary variables are assumed to be known either for all population units or through known domain totals. This information is used for participation modeling, domain calibration, and the construction of area-level covariates.

Let $\bx_i$ be a vector of unit-level covariates used to model membership in the non-probability sample. We use a logistic participation model
\begin{equation}\label{eq:propensity_model}
\logit(\pi_i)=\bx_i^\T\bgamma,
\qquad
\pi_i=P(\delta_i=1\mid \bx_i).
\end{equation}
The fitted probabilities are denoted by $\hat\pi_i$, and the raw IPW weights are
\begin{equation*}
w_i^{(0)}=\hat\pi_i^{-1},
\qquad i\in S.
\end{equation*}
Domain effects are not included directly in \eqref{eq:propensity_model}; domain information enters through the calibration equations below. This keeps the participation model parsimonious and avoids unstable domain-effect fits in small domains.

Let $\bx_i^{\rm cal}$ be the vector of calibration variables. These variables may be selected from the covariates in $\bx_i$ or from other auxiliary variables whose domain totals are known. Define $\bX_d^{\rm cal}=\sum_{i\in U_d}\bx_i^{\rm cal}$. Following the calibration principle of \citet{DevilleSarndal1992}, the calibrated weights are chosen within each domain by a quadratic distance from the raw IPW weights. Specifically, $w_i$, $i\in S_d$, solve
\begin{equation}\label{eq:calibration_problem}
\min_{\{w_i:i\in S_d\}}
\sum_{i\in S_d}
\frac{(w_i-w_i^{(0)})^2}{w_i^{(0)}}
\quad \text{subject to}\quad
\sum_{i\in S_d} w_i\bx_i^{\rm cal}=\bX_d^{\rm cal},
\qquad d=1,\ldots,D.
\end{equation}
The constraint in this problem is the domain calibration equation. When $\bx_i^{\rm cal}=1$, it calibrates to the domain population size, $\sum_{i\in S_d}w_i=N_d$. When $\bx_i^{\rm cal}$ also contains auxiliary variables, it calibrates to their known domain totals. Thus, the calibration information set can range from domain size only to a richer vector of known auxiliary totals. In applications, this choice is part of the estimation specification; adding auxiliary totals is not automatically beneficial in finite domains.

The calibrated IPW domain estimator is
\begin{equation}\label{eq:ipw_estimator}
\hat Y_d^{\rm IPW}=\sum_{i\in S_d}w_i y_i.
\end{equation}
This is the area-level input to the FH stage. We call it a calibrated IPW domain estimator rather than a design-based direct estimator, because its validity depends on the fitted participation model and the calibration information. Its variance input, denoted by $\widehat V_d$, is estimated by a generalized multiplier bootstrap that refits the participation model and repeats the calibration in each replicate; the procedure is given in Appendix~\ref{app:direct_bootstrap}.

\section{Domain diagnostics and reliability score}\label{sec:diagnostics}

The method uses three diagnostics to summarize the quality of the calibrated IPW information in each domain. The first diagnostic is the domain coverage rate,
\begin{equation*}
c_d=\frac{n_d}{N_d}.
\end{equation*}
Coverage records how much of the domain population is represented before the observed units are reweighted to represent the whole domain.

The second diagnostic is the coefficient of variation of the calibrated weights in the domain,
\begin{equation}\label{eq:weight_cv}
\CV_{w,d}=
\frac{\left\{n_d^{-1}\sum_{i\in S_d}(w_i-\bar w_d)^2\right\}^{1/2}}
{\bar w_d},
\qquad
\bar w_d=n_d^{-1}\sum_{i\in S_d}w_i.
\end{equation}
Large values indicate that the IPW estimator is strongly affected by a small number of weighted units. The usual effective sample size is
\begin{equation*}
n_d^{\rm eff}=\frac{(\sum_{i\in S_d}w_i)^2}{\sum_{i\in S_d}w_i^2}.
\end{equation*}
Using \eqref{eq:weight_cv}, this can be written as $n_d^{\rm eff}=n_d/(1+\CV_{w,d}^2)$. Hence, the corresponding effective coverage is
\begin{equation*}
c_d^{\rm eff}=\frac{n_d^{\rm eff}}{N_d}=\frac{c_d}{1+\CV_{w,d}^2}.
\end{equation*}
Thus, effective coverage is not an additional diagnostic dimension; it is determined by the first two diagnostics, $c_d$ and $\CV_{w,d}$.

The third diagnostic measures auxiliary balance. Let $\bh_i$ be the vector of variables used for this assessment, transformed componentwise to population mean zero and population variance one. That is, if $h_{ij}^{0}$ is an unstandardized auxiliary variable, then
\begin{equation*}
h_{ij}=\frac{h_{ij}^{0}-\bar h_{Uj}^{0}}{s_{Uj}^{0}},
\end{equation*}
where $\bar h_{Uj}^{0}=N^{-1}\sum_{i\in U}h_{ij}^{0}$ is the finite-population mean and $s_{Uj}^{0}$ is the corresponding finite-population standard deviation. Define the population and weighted sample means in the domain $d$ by
\begin{equation*}
\bar{\bh}_{U,d}=N_d^{-1}\sum_{i\in U_d}\bh_i,
\qquad
\bar{\bh}_{w,d}=\left(\sum_{i\in S_d}w_i\right)^{-1}\sum_{i\in S_d}w_i\bh_i.
\end{equation*}
The scalar balance discrepancy is the Euclidean distance
\begin{equation*}
B_d=
\left[
\sum_{j=1}^{p_h}
\left(\bar h_{w,d,j}-\bar h_{U,d,j}\right)^2
\right]^{1/2},
\end{equation*}
where $p_h$ is the dimension of $\bh_i$. A larger $B_d$ means that the calibrated weighted sample remains less aligned with known population auxiliary information.

The diagnostics are combined only for domains with enough information for the diagnostics to be meaningful. Let
\begin{equation*}
\mathcal E=\{d:N_d\ge N_0,\ n_d\ge n_0,\ \widehat V_d<\infty\}
\end{equation*}
be the eligible set, where $N_0$ and $n_0$ are fixed minimum domain-size constants. Let $m=|\mathcal E|$. For a diagnostic $q_d$, let $\mathcal R_{\mathcal E}(q_d)$ be its mid-rank among the values $\{q_{d'}:d'\in\mathcal E\}$, with rank one assigned to the smallest value. For diagnostics where larger values are better, define
\begin{equation*}
R_d^+(q)=\frac{\mathcal R_{\mathcal E}(q_d)-1}{m-1},
\qquad d\in\mathcal E.
\end{equation*}
For diagnostics where smaller values are better, define
\begin{equation*}
R_d^-(q)=\frac{m-\mathcal R_{\mathcal E}(q_d)}{m-1},
\qquad d\in\mathcal E.
\end{equation*}
Both scores lie in $[0,1]$, with larger values always corresponding to better diagnostics. The reliability score is
\begin{equation*}
Q_d=\frac{1}{3}\left\{R_d^+(c)+R_d^-(\CV_w)+R_d^-(B)\right\},
\qquad d\in\mathcal E.
\end{equation*}
Domains outside $\mathcal E$ are assigned the lowest reliability level when the mixture component probabilities are formed below. This construction keeps the diagnostics interpretable, puts them on a common scale, and gives equal weight to the three diagnostic dimensions.

\section{Diagnostics-guided mixture variance-inflated Fay--Herriot estimation}\label{sec:model}

Let $\bz_d$ be a vector of known domain-level covariates, including an intercept unless stated otherwise, and write
\begin{equation*}
m_d(\bbeta)=\bz_d^\T\bbeta.
\end{equation*}
The standard FH model for the calibrated IPW input is
\begin{equation*}
\hat Y_d^{\rm IPW}=\theta_d+e_d,
\qquad
\theta_d=m_d(\bbeta)+u_d,
\qquad
u_d\sim N(0,A),
\qquad
e_d\sim N(0,\widehat V_d),
\end{equation*}
with independent random effects and input errors. In the non-probability setting, however, the variance estimate supplied to the FH model may not capture all reliability concerns. A domain can have a small variance estimate while still being weakly supported because of low coverage, unstable weights, or poor balance. Let
\begin{equation*}
\mathcal F=\{d:n_d>0,\ \widehat V_d<\infty\}
\end{equation*}
denote the set of domains used in the area-level fitting step.

To represent this additional uncertainty, introduce a latent component indicator $\xi_d\in\{0,1\}$, where $\xi_d=1$ denotes the ordinary component and $\xi_d=0$ denotes the variance-inflated component. Conditional on $\theta_d$ and $\xi_d$, define
\begin{equation}\label{eq:mixture_observation}
\hat Y_d^{\rm IPW}\mid \theta_d,\xi_d
\sim
N\{\theta_d,
       \widehat V_d+(1-\xi_d)\tau\},
\qquad
\tau\ge 0.
\end{equation}
Thus, the inflated component carries an additional variance term $\tau$. The prior probability of the ordinary component is determined by the diagnostic score. First, rescale $Q_d$ on the eligible domains by
\begin{equation*}
\widetilde Q_d=
\frac{Q_d-Q_{\min}}{Q_{\max}-Q_{\min}},
\qquad
Q_{\min}=\min_{d\in\mathcal E}Q_d,
\quad
Q_{\max}=\max_{d\in\mathcal E}Q_d,
\end{equation*}
with $\widetilde Q_d=1/2$ if $Q_{\max}=Q_{\min}$. Then set
\begin{equation*}
\pi_d=P(\xi_d=1)=
\begin{cases}
\widetilde Q_d, & d\in\mathcal E,\\
0, & d\notin\mathcal E.
\end{cases}
\end{equation*}
The ordinary FH model is recovered when $\tau=0$ or when $\pi_d=1$ for all domains.

Combining \eqref{eq:mixture_observation} with the FH linking model gives the two marginal component densities
\begin{align*}
\hat Y_d^{\rm IPW}\mid \xi_d=1
&\sim N\{m_d(\bbeta),A+\widehat V_d\},\\
\hat Y_d^{\rm IPW}\mid \xi_d=0
&\sim N\{m_d(\bbeta),A+\widehat V_d+\tau\}.
\end{align*}
Let $\varphi(y;\mu,\sigma^2)$ denote the normal density with mean $\mu$ and variance $\sigma^2$ evaluated at $y$.

The estimator is computed by an iterative FH algorithm. Initialization uses the highest-ranked eligible domains only to obtain starting values, not to define a separate estimator. Let $K_0$ be larger than the number of regression coefficients in $m_d(\bbeta)$, and let $\mathcal S_0$ be the $K_0$ domains in $\mathcal E$ with the largest $Q_d$. A standard FH fit on $\mathcal S_0$ gives initial values $\bbeta^{(0)}$ and $A^{(0)}$. The initial inflation scale is taken as a nonnegative excess-residual scale, for example,
\begin{equation*}
\tau^{(0)}=
\operatorname{median}_{d\in\mathcal E\setminus\mathcal S_0}
\left[
\max\{0,
(\hat Y_d^{\rm IPW}-m_d(\bbeta^{(0)}))^2-\widehat V_d-A^{(0)}\}
\right],
\end{equation*}
with the convention $\tau^{(0)}=0$ if $\mathcal E\setminus\mathcal S_0$ is empty.

Suppose that at iteration $t-1$ the current values are $\bbeta^{(t-1)}$, $A^{(t-1)}$ and $\tau^{(t-1)}$. The current ordinary and inflated marginal densities are
\begin{align*}
g_d^{(t-1)}&=\varphi\{\hat Y_d^{\rm IPW};m_d(\bbeta^{(t-1)}),A^{(t-1)}+\widehat V_d\},\\
b_d^{(t-1)}&=\varphi\{\hat Y_d^{\rm IPW};m_d(\bbeta^{(t-1)}),A^{(t-1)}+\widehat V_d+\tau^{(t-1)}\}.
\end{align*}
Bayes' rule gives the posterior probability of the ordinary component,
\begin{equation*}
p_d^{(t)}=
\frac{\pi_d g_d^{(t-1)}}
{\pi_d g_d^{(t-1)}+(1-\pi_d)b_d^{(t-1)}}.
\end{equation*}
Given $p_d^{(t)}$, form the posterior-averaged input variance
\begin{equation*}
\widehat V_d^{\star(t)}=\widehat V_d+\{1-p_d^{(t)}\}\tau^{(t-1)}.
\end{equation*}
Fit the standard FH model on $\mathcal F$ with input variance estimates $\widehat V_d^{\star(t)}$ to update $\bbeta^{(t)}$ and $A^{(t)}$. The inflation parameter is updated on the same fitting set by the posterior-weighted excess-residual moment equation
\begin{equation*}
\tau_{\rm raw}^{(t)}=
\frac{
\sum_{d\in\mathcal F}\{1-p_d^{(t)}\}
\max\{0,(\hat Y_d^{\rm IPW}-m_d(\bbeta^{(t)}))^2-\widehat V_d-A^{(t)}\}}
{\sum_{d\in\mathcal F}\{1-p_d^{(t)}\}},
\end{equation*}
whenever the denominator is positive; otherwise $\tau_{\rm raw}^{(t)}=\tau^{(t-1)}$. A damped update
\begin{equation*}
\tau^{(t)}=\lambda_\tau\tau^{(t-1)}+(1-\lambda_\tau)\tau_{\rm raw}^{(t)},
\qquad 0\le \lambda_\tau<1,
\end{equation*}
is used to stabilize the iteration; in the empirical analyses below, we use $\lambda_\tau=0.60$. The iteration stops when relative changes in $\bbeta$, $A$, $\tau$, and the posterior probabilities are below a fixed tolerance.

At convergence, let $\hat\bbeta$, $\hat A$, $\hat\tau$ and $\hat p_d$ denote the final values. The component-specific shrinkage factors are
\begin{equation*}
\gamma_{d1}=\frac{\hat A}{\hat A+\widehat V_d},
\qquad
\gamma_{d0}=\frac{\hat A}{\hat A+\widehat V_d+\hat\tau}.
\end{equation*}
The component-specific predictors are
\begin{equation*}
\hat\theta_{dr}=m_d(\hat\bbeta)+\gamma_{dr}\{\hat Y_d^{\rm IPW}-m_d(\hat\bbeta)\},
\qquad r=0,1.
\end{equation*}
The proposed domain predictor is the posterior average
\begin{equation}\label{eq:proposed_predictor}
\hat Y_d^{\rm VI}=\hat p_d\hat\theta_{d1}+(1-\hat p_d)\hat\theta_{d0}.
\end{equation}
Thus, domains with stronger diagnostics and smaller residual evidence for inflation retain more of their calibrated IPW input, whereas domains with weaker diagnostics are smoothed more strongly toward the FH mean.

For inference on $\hat Y_d^{\rm VI}$, we use a full-pipeline bootstrap MSE estimator. The bootstrap recomputes the participation model, calibration, IPW inputs, variance estimates, diagnostics, and the variance-inflated FH fit in each full-pipeline replicate. The resulting MSE estimator for $\hat Y_d^{\rm VI}$ is given in \eqref{eq:full_boot_mse} and described in Appendix~\ref{app:full_bootstrap}.

\section{Truth-known business-turnover validation}\label{sec:services}

\subsection{Data and validation design}

The study is based on data provided by the State Data Agency (Statistics Lithuania). The validation uses a finite pseudo-real population of Lithuanian service-activity enterprises; below, we simply refer to them as business enterprises. The population contains $N=2\,839$ enterprises divided into $D=23$ economic activity domains. The study variable is monthly turnover. Auxiliary information is available for all population units from administrative registers; in our analysis, the main auxiliary variables are previous-year annual turnover and employment.

The non-probability sample is obtained from administrative value-added-tax records and contains $n=2\,232$ enterprises. Since small enterprises are often not value-added-tax payers, the administrative source over-represents large units and is heterogeneous relative to its complement. This makes the data useful for evaluating methods based on non-probability IPW inputs. The full pseudo-population values of monthly turnover are known and are used here only for validation; model fitting treats the target variable as observed only for units in the administrative non-probability source.

We compare three area-level estimators. The first is the calibrated IPW estimator \eqref{eq:ipw_estimator}. The second is the standard FH empirical best linear unbiased predictor fitted to the same calibrated IPW inputs and variance estimates, but without diagnostics-based variance inflation. The third is the proposed variance-inflated FH estimator \eqref{eq:proposed_predictor}. Three calibration information sets are considered: calibration to domain population sizes; calibration additionally to the annual-turnover domain total; and calibration additionally to annual-turnover and employment totals. The latter two settings examine how the method behaves when richer auxiliary calibration information is available; they are not assumed to be more reliable a priori. The eligible set used in the reliability construction is formed with $N_0=10$ and $n_0=5$.

For each method and calibration information set, we compute the aggregate relative error, the domain root mean squared error (RMSE), and the domain mean absolute percentage error (MAPE), using the known domain totals $Y_d$ as truth. For a generic estimator $\hat Y_d$, these measures are
\begin{align*}
\mathrm{RE}_{\rm agg}
&=100\,\frac{\sum_{d=1}^D \hat Y_d-\sum_{d=1}^D Y_d}{\sum_{d=1}^D Y_d},\\
\mathrm{RMSE}_{\rm dom}
&=\left\{D^{-1}\sum_{d=1}^D(\hat Y_d-Y_d)^2\right\}^{1/2},\\
\mathrm{MAPE}_{\rm dom}
&=100\,D^{-1}\sum_{d=1}^D\left|\frac{\hat Y_d-Y_d}{Y_d}\right|.
\end{align*}

\subsection{Validation results}

Table~\ref{tab:services_validation} gives the main validation results. The standard FH benchmark gives only a modest improvement over the calibrated IPW input. In contrast, the proposed variance-inflated estimator improves substantially over both calibrated IPW and the standard FH benchmark. Under calibration to domain population sizes, the aggregate relative error decreases from 14.56\% for calibrated IPW and 12.83\% for standard FH to 2.67\% for the proposed estimator. The domain RMSE decreases from 7.87 million and 7.59 million to 1.31 million, and the MAPE decreases from 61.22\% and 57.68\% to 13.39\%.

\begin{table}[H]
\centering
\caption{Truth-known business-turnover validation. Errors are computed against the known domain totals. RMSE is reported in millions of turnover units.}
\label{tab:services_validation}
\small
\begin{tabular}{@{}p{0.34\textwidth}lrrr@{}}
\toprule
Information set & Method &
\shortstack{Aggregate\\rel. error} &
\shortstack{Domain\\RMSE} &
MAPE\\
\midrule
Domain size & Calibrated IPW & 14.56\% & 7.87 & 61.22\%\\
Domain size & Standard FH & 12.83\% & 7.59 & 57.68\%\\
Domain size & Proposed VI-FH & \textbf{2.67\%} & \textbf{1.31} & \textbf{13.39\%}\\
\addlinespace
Domain size + auxiliary turnover & Calibrated IPW & 16.58\% & 7.27 & 38.17\%\\
Domain size + auxiliary turnover & Standard FH & 16.00\% & 7.01 & 38.41\%\\
Domain size + auxiliary turnover & Proposed VI-FH & \textbf{6.10\%} & \textbf{2.29} & \textbf{34.07\%}\\
\addlinespace
Domain size + auxiliary turnover + employment & Calibrated IPW & 16.84\% & 7.28 & 38.36\%\\
Domain size + auxiliary turnover + employment & Standard FH & 16.11\% & 7.00 & 38.46\%\\
Domain size + auxiliary turnover + employment & Proposed VI-FH & \textbf{5.95\%} & \textbf{2.29} & \textbf{35.75\%}\\
\bottomrule
\end{tabular}
\end{table}

The comparison across calibration information sets is also informative. In this pseudo-population, calibration only to domain size gives the smallest errors for the proposed estimator. Adding auxiliary totals changes both the calibration constraints and the resulting weight distribution within domains. The calibration information set should therefore be treated as part of the statistical specification, rather than as a monotone path to improved small-area estimation.

\begin{figure}[H]
\centering
\includegraphics[width=0.92\textwidth]{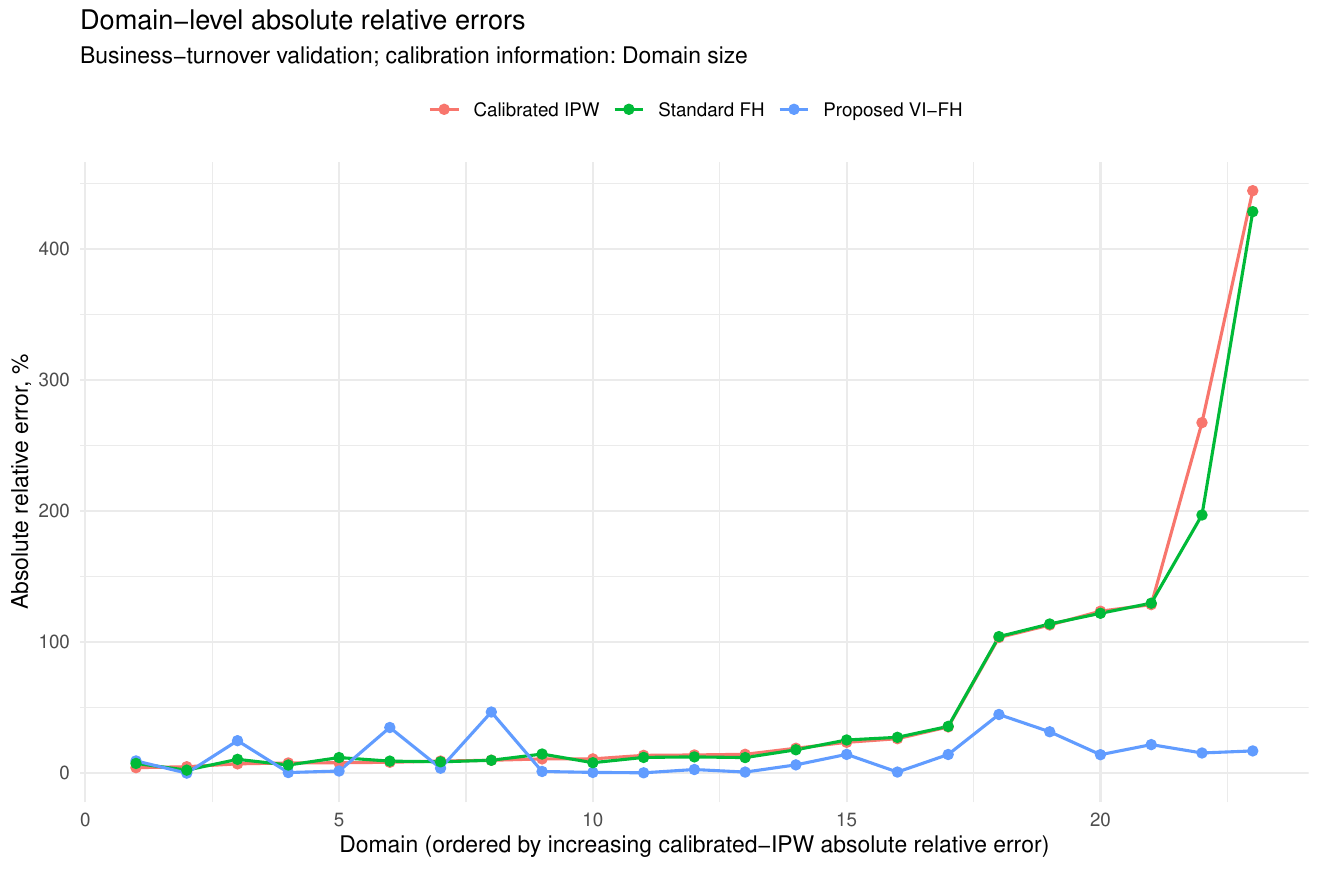}
\caption{Domain-level absolute relative errors for the domain-size calibration setting. Domains are ordered by increasing calibrated-IPW absolute relative error.}
\label{fig:services_are}
\end{figure}

\begin{figure}[H]
\centering
\includegraphics[width=0.92\textwidth]{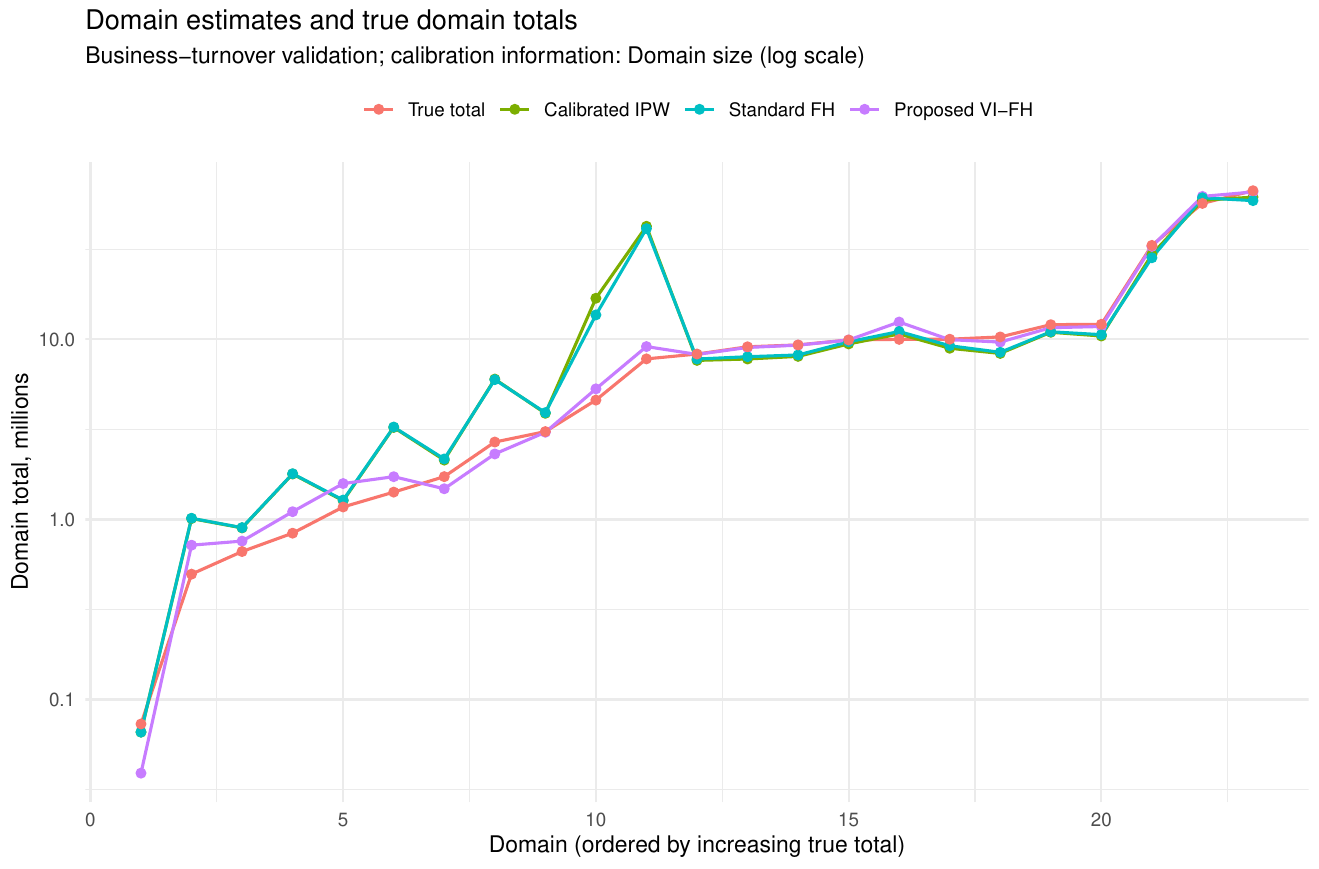}
\caption{Domain estimates and true domain totals for the domain-size calibration setting. Domains are ordered by increasing true total; the vertical axis is logarithmic.}
\label{fig:services_estimates_log}
\end{figure}

Figures~\ref{fig:services_are} and~\ref{fig:services_estimates_log} show the domain-level behavior for the domain-size calibration setting. Figure~\ref{fig:services_are} displays the absolute relative errors by domain, with domains ordered by increasing calibrated-IPW error. It shows that the standard FH benchmark largely follows the calibrated IPW input in the most problematic domains, whereas the proposed estimator substantially reduces the largest domain errors. Figure~\ref{fig:services_estimates_log} compares the estimated and true domain totals on a logarithmic scale, showing that the proposed estimator follows the known domain totals more closely across the ordered domains.

\subsection{Validation of the diagnostics}

The truth-known setting also lets us check whether the diagnostics are empirically informative. Let
\begin{equation}\label{eq:ARE_direct}
\mathrm{ARE}^{\rm IPW}_d=\frac{|\hat Y_d^{\rm IPW}-Y_d|}{Y_d}
\end{equation}
be the absolute relative error of the calibrated IPW estimator. Since the diagnostics are designed to describe the reliability of the IPW input before FH smoothing, \eqref{eq:ARE_direct} is the natural validation target.

Table~\ref{tab:diagnostic_correlations} reports correlations between $\mathrm{ARE}^{\rm IPW}_d$ and three diagnostic quantities. The signs are consistent with the intended interpretation. Higher coverage and a higher reliability score are associated with smaller actual IPW errors, while a larger auxiliary imbalance is associated with larger IPW errors. The correlations are computed over the $20$ domains for which the reliability score is defined.

\begin{table}[H]
\centering
\caption{Correlation between calibrated-IPW absolute relative error and diagnostic quantities in the truth-known business validation.}
\label{tab:diagnostic_correlations}
\small
\begin{tabular}{@{}lrrr@{}}
\toprule
Information set &
\shortstack{cor($Q_d$,\\ARE$_d^{\rm IPW}$)} &
\shortstack{cor($c_d$,\\ARE$_d^{\rm IPW}$)} &
\shortstack{cor($B_d$,\\ARE$_d^{\rm IPW}$)}\\
\midrule
Domain size & -0.599 & -0.900 & 0.790\\
Domain size + auxiliary turnover & -0.468 & -0.764 & 0.749\\
Domain size + auxiliary turnover + employment & -0.453 & -0.753 & 0.754\\
\bottomrule
\end{tabular}
\end{table}

These results validate the role of the diagnostics independently of the smoothing step. The diagnostics do not merely provide a plausible ranking; in this population, they are empirically related to the actual error of the calibrated IPW estimator.

\section{Discussion}\label{sec:discussion}

The proposed diagnostics-guided variance-inflated FH estimator addresses a practical weakness of standard area-level smoothing in non-probability settings. Calibrated IPW domain estimators can be useful, but their quality varies across domains. Treating all input variances as if they fully captured domain reliability can lead to excessive trust in domains with weak coverage, unstable weights, or poor auxiliary balance. The proposed mixture variance-inflation mechanism uses these diagnostics to moderate the influence of less reliable IPW inputs.

The business-turnover validation gives empirical evidence for the method. In a truth-known pseudo-population, the proposed estimator reduced aggregate and domain-level errors substantially relative to calibrated IPW estimation. The validation also showed that the diagnostics have the intended relationship with the actual IPW error. The diagnostics are not only interpretable; they are empirically informative in a population where the truth is known.

The validation results also show that auxiliary calibration should be interpreted as a modeling specification rather than as an automatic improvement. Richer calibration information is not automatically better. Calibration to additional auxiliary totals can improve some balance measures, but it also changes the constraints imposed on the domain weights. In the business data, the best overall proposed-estimator performance was obtained by calibration to domain population sizes. This does not imply that auxiliary totals should be avoided; rather, it shows that the calibration information set is a statistical choice that should be examined together with diagnostics and estimation error.

The method does not remove the need for assumptions. It relies on observed auxiliary variables, a fitted participation model, domain calibration, and an area-level mean structure. It cannot identify arbitrary unmeasured selection bias without information. Its purpose is more specific: to prevent the area-level model from treating all non-probability IPW inputs as equally reliable after conditioning only on their variance estimates. The empirical study is deliberately limited to one truth-known business pseudo-population. It should therefore be read as targeted evidence on the proposed variance-inflation mechanism, not as an exhaustive assessment of all possible non-probability selection mechanisms.

\section*{Acknowledgements}
This project has received funding from the Research Council of Lithuania (LMTLT), agreement No~S-MIP-25-10.

\appendix

\section{Generalized multiplier bootstrap for calibrated IPW inputs}\label{app:direct_bootstrap}

This appendix gives the bootstrap procedure used to evaluate the variance estimate $\widehat V_d$ for the calibrated IPW domain estimators. Let $B_1$ be the number of bootstrap replicates. For replicate $b=1,\ldots,B_1$, generate independent positive multiplier weights $\xi_i^{(b)}$, $i\in U$, such that
\begin{equation*}
\E\{\xi_i^{(b)}\}=1,
\qquad
\Var\{\xi_i^{(b)}\}=1.
\end{equation*}
In the empirical computations, we use exponential multipliers, $\xi_i^{(b)}\sim\mathrm{Exp}(1)$. The procedure is related to generalized bootstrap ideas for survey estimation \citep{BeaumontPatak2012}, adapted here to perturb the fitted participation model and the calibration step.

For each bootstrap replicate $b$, the following steps are carried out.
\begin{enumerate}[topsep=3pt,itemsep=2pt]
\item Maximize the multiplier-weighted logistic log-likelihood
\begin{equation*}
\ell^{*(b)}(\bgamma)=
\sum_{i\in U}\xi_i^{(b)}
\left[
\delta_i\log \pi_i(\bgamma)+
(1-\delta_i)\log\{1-\pi_i(\bgamma)\}
\right],
\end{equation*}
where $\logit\{\pi_i(\bgamma)\}=\bx_i^\T\bgamma$. Denote the resulting fitted probabilities by $\hat\pi_i^{*(b)}$.

\item Construct raw bootstrap IPW weights
\begin{equation*}
w_i^{(0)*(b)}=\{\hat\pi_i^{*(b)}\}^{-1},
\qquad i\in S.
\end{equation*}

\item Apply the same type of domain calibration as in \eqref{eq:calibration_problem}. In replicate $b$, the bootstrap calibrated weights are obtained by solving
\begin{equation*}
\min_{\{w_i:i\in S_d\}}
\sum_{i\in S_d}\xi_i^{(b)}
\frac{\{w_i-w_i^{(0)*(b)}\}^2}{w_i^{(0)*(b)}}
\quad \text{subject to}\quad
\sum_{i\in S_d}\xi_i^{(b)}w_i\bx_i^{\rm cal}
=
\bX_d^{\rm cal},
\qquad d=1,\ldots,D.
\end{equation*}
Set $w_i^{*(b)}$ equal to this solution.

\item Recompute the calibrated IPW domain total
\begin{equation*}
\hat Y_d^{{\rm IPW},*(b)}=
\sum_{i\in S_d}\xi_i^{(b)}w_i^{*(b)}y_i.
\end{equation*}
\end{enumerate}
The variance estimate input for the FH stage is
\begin{equation*}
\widehat V_d=
\frac{1}{B_1-1}
\sum_{b=1}^{B_1}
\left(\hat Y_d^{{\rm IPW},*(b)}-\bar Y_d^{{\rm IPW},*}\right)^2, \qquad \bar Y_d^{{\rm IPW},*}=B_1^{-1}\sum_{b=1}^{B_1}\hat Y_d^{{\rm IPW},*(b)}.
\end{equation*}

\section{Full-pipeline bootstrap MSE estimator}\label{app:full_bootstrap}

This appendix gives the bootstrap procedure used to estimate the MSE of the final variance-inflated predictor. It is distinct from the multiplier bootstrap in Appendix~\ref{app:direct_bootstrap}: the latter supplies the input variance estimates $\widehat V_d$, whereas the present procedure propagates uncertainty through the complete estimation pipeline.

Let $B_2$ be the number of full-pipeline bootstrap replicates. In replicate $b$, a bootstrap finite population is generated within each domain from the observed donor units. For a domain $d$, draw copy counts $M_i^{*(b)}$, $i\in S_d$, from
\begin{equation*}
(M_i^{*(b)}:i\in S_d)
\sim
\mathrm{Multinomial}
\left(N_d,
\left\{\frac{w_i}{\sum_{j\in S_d}w_j}:i\in S_d\right\}
\right),
\end{equation*}
where $w_i$ are the calibrated IPW weights from the original fit. The bootstrap population total is
\begin{equation*}
Y_d^{*(b)}=\sum_{i\in S_d}M_i^{*(b)}y_i .
\end{equation*}

A bootstrap non-probability sample is then generated from this bootstrap population. Conditional on the copy counts, selected-copy counts are drawn as
\begin{equation*}
L_i^{*(b)}\mid M_i^{*(b)}
\sim
\mathrm{Binomial}\{M_i^{*(b)},\hat\pi_i\},
\qquad i\in S_d,
\end{equation*}
where $\hat\pi_i$ is the fitted participation probability from the original analysis. Equivalently, each generated copy inherits the donor's covariates, outcome, and fitted participation probability, and is then independently included in the bootstrap non-probability sample with that inherited probability.

On the bootstrap sample, the whole estimator is recomputed: the participation model is refitted, the weights are recalibrated, the calibrated IPW totals and their variance inputs are recomputed, the diagnostics and reliability score are reconstructed, and the variance-inflated FH estimator is refitted. Denote the resulting bootstrap predictor by $\hat Y_d^{{\rm VI},*(b)}$.

The full-pipeline MSE estimator for $\hat Y_d^{\rm VI}$ is
\begin{equation}\label{eq:full_boot_mse}
\widehat{\mathrm{MSE}}(\hat Y_d^{\rm VI})
=
\frac{1}{B_2}
\sum_{b=1}^{B_2}
\left(\hat Y_d^{{\rm VI},*(b)}-Y_d^{*(b)}\right)^2 .
\end{equation}


\begin{thebibliography}{99}

\bibitem[Beaumont and Patak(2012)]{BeaumontPatak2012}
Beaumont, J.-F. and Patak, Z. (2012).
On the generalized bootstrap for sample surveys with special attention to Poisson sampling.
\emph{International Statistical Review}, 80, 127--148.

\bibitem[Chakraborty et~al.(2016)]{ChakrabortyDattaMandal2016}
Chakraborty, A., Datta, G. S. and Mandal, A. (2016).
A two-component normal mixture alternative to the Fay--Herriot model.
\emph{Statistics in Transition new series}, 17, 67--90.

\bibitem[Chambers et~al.(2014)]{ChambersEtAl2014}
Chambers, R., Chandra, H., Salvati, N. and Tzavidis, N. (2014).
Outlier robust small area estimation.
\emph{Journal of the Royal Statistical Society: Series B}, 76, 47--69.

\bibitem[Chen et~al.(2020)]{ChenLiWu2020}
Chen, Y., Li, P. and Wu, C. (2020).
Doubly robust inference with nonprobability survey samples.
\emph{Journal of the American Statistical Association}, 115, 2011--2021.

\bibitem[Choi et~al.(2021)]{ChoiNandramKim2021}
Choi, S., Nandram, B. and Kim, D. (2021).
Bayesian predictive inference of small area proportions under selection bias.
\emph{Survey Methodology}, 47, 91--122.

\bibitem[Datta and Mandal(2015)]{DattaMandal2015}
Datta, G. S. and Mandal, A. (2015).
Small area estimation with uncertain random effects.
\emph{Journal of the American Statistical Association}, 110, 1735--1744.

\bibitem[Deville and S{\"a}rndal(1992)]{DevilleSarndal1992}
Deville, J.-C. and S{\"a}rndal, C.-E. (1992).
Calibration estimators in survey sampling.
\emph{Journal of the American Statistical Association}, 87, 376--382.

\bibitem[Elliott and Valliant(2017)]{ElliottValliant2017}
Elliott, M. R. and Valliant, R. (2017).
Inference for nonprobability samples.
\emph{Statistical Science}, 32, 249--264.

\bibitem[Fay and Herriot(1979)]{FayHerriot1979}
Fay, R. E. and Herriot, R. A. (1979).
Estimates of income for small places: an application of James--Stein procedures to census data.
\emph{Journal of the American Statistical Association}, 74, 269--277.

\bibitem[Jiang and Rao(2020)]{JiangRao2020}
Jiang, J. and Rao, J. S. (2020).
Robust small area estimation: an overview.
\emph{Annual Review of Statistics and Its Application}, 7, 337--360.

\bibitem[Kim and Tam(2021)]{KimTam2021}
Kim, J. K. and Tam, S.-M. (2021).
Data integration by combining big data and survey sample data for finite population inference.
\emph{International Statistical Review}, 89, 382--401.

\bibitem[Liu et~al.(2026)]{LiuEtAl2026}
Liu, A.-C., Scholtus, S., Van~Deun, K. and de~Waal, T. (2026).
Domain estimation from weighted nonprobability samples.
\emph{Statistical Journal of the IAOS}. Online first.

\bibitem[Lohr(2021)]{Lohr2021}
Lohr, S. L. (2021).
Multiple-frame surveys for a multiple-data-source world.
\emph{Survey Methodology}, 47, 229--263.

\bibitem[Marchetti et~al.(2015)]{MarchettiEtAl2015}
Marchetti, S., Giusti, C., Pratesi, M., Salvati, N., Giannotti, F., Pedreschi, D., Rinzivillo, S., Pappalardo, L. and Gabrielli, L. (2015).
Small area model-based estimators using big data sources.
\emph{Journal of Official Statistics}, 31, 263--281.

\bibitem[Rao(2021)]{Rao2021}
Rao, J. N. K. (2021).
On making valid inferences by integrating data from surveys and other sources.
\emph{Sankhya B}, 83, 242--272.

\bibitem[Rao and Molina(2015)]{RaoMolina2015}
Rao, J. N. K. and Molina, I. (2015).
\emph{Small Area Estimation}, 2nd ed.
Wiley, Hoboken, NJ.

\bibitem[Savitsky et~al.(2023)]{SavitskyEtAl2023}
Savitsky, T. D., Williams, M. R., Gershunskaya, J. and Beresovsky, V. (2023).
Methods for combining probability and nonprobability samples under unknown overlaps.
\emph{Statistics in Transition new series}, 24, 1--34.

\bibitem[Savitsky et~al.(2025)]{SavitskyEtAl2025}
Savitsky, T. D., Williams, M. R., Beresovsky, V. and Gershunskaya, J. (2025).
Thresholding nonprobability units in combined data for efficient domain estimation.
\emph{Statistics in Transition new series}, 26, 1--19.

\bibitem[Schirripa~Spagnolo et~al.(2025)]{SchirripaSpagnoloEtAl2025}
Schirripa~Spagnolo, F., Bertarelli, G., Summa, D., Scannapieco, M., Pratesi, M., Marchetti, S. and Salvati, N. (2025).
Inference for big data assisted by small area methods: an application on sustainable development goals sensitivity in Italian enterprises.
\emph{Journal of the Royal Statistical Society Series A: Statistics in Society}, 188, 27--45.

\bibitem[Sinha and Rao(2009)]{SinhaRao2009}
Sinha, S. K. and Rao, J. N. K. (2009).
Robust small area estimation.
\emph{Canadian Journal of Statistics}, 37, 381--399.

\bibitem[\v{S}levinskas et~al.(2026)]{SlevinskasEtAl2026}
\v{S}levinskas, D., Burakauskait{\.e}, I. and \v{C}iginas, A. (2026).
Small area estimation using incomplete auxiliary information.
\emph{arXiv preprint} arXiv:2602.12845.

\bibitem[Villalobos-Aliste et~al.(2025)]{VillalobosAlisteEtAl2025}
Villalobos-Aliste, S. F., Scholtus, S. and de~Waal, T. (2025).
Combining probability and nonprobability samples on an aggregated level.
\emph{Journal of Official Statistics}, 41, 619--648.

\bibitem[Wu(2022)]{Wu2022}
Wu, C. (2022).
Statistical inference with non-probability survey samples.
\emph{Survey Methodology}, 48, 283--311.

\bibitem[Yang and Kim(2020)]{YangKim2020}
Yang, S. and Kim, J. K. (2020).
Statistical data integration in survey sampling: a review.
\emph{Japanese Journal of Statistics and Data Science}, 3, 625--650.

\bibitem[Yang et~al.(2021)]{YangKimHwang2021}
Yang, S., Kim, J. K. and Hwang, Y. (2021).
Integration of data from probability surveys and big found data for finite population inference using mass imputation.
\emph{Survey Methodology}, 47, 29--58.

\end{thebibliography}
\end{document}